\documentclass[journal]{IEEEtran}
\ifCLASSINFOpdf
   \usepackage[pdftex]{graphicx}
\else
\fi
\usepackage{color,soul}

\ifCLASSOPTIONcompsoc
  \usepackage[caption=false,font=normalsize,labelfont=sf,textfont=sf]{subfig}
\else
  \usepackage[caption=false,font=footnotesize]{subfig}
\fi
\begin{document}
%
\title{Seamless roaming and guaranteed communication using a synchronized single-hop multi-gateway 802.15.4e TSCH network\footnote{This is a short preprint version of a paper with the same title that is in the process of reviewing at Ad-Hoc Networks Journal.}}
%
%
%

\author{\IEEEauthorblockN{J. Haxhibeqiri\IEEEauthorrefmark{1}~(jetmir.haxhibeqiri@ugent.be),
        A. Karaagac\IEEEauthorrefmark{1},
        I. Moerman\IEEEauthorrefmark{1},
        and~J. Hoebeke\IEEEauthorrefmark{1}\\
}
\IEEEauthorblockN{
\IEEEauthorrefmark{1} IDLab, Ghent University -- imec, Ghent, Belgium.
}}

\maketitle

\begin{abstract}
Industrial wireless sensor networks (WSNs) are being used to improve the efficiency, productivity and safety of industrial processes. An open standard that is commonly used in such cases is IEEE 802.15.4e. Its TSCH mode employs a time synchronized based MAC scheme together with channel hopping to alleviate the impact of channel fading. Until now, most of the industrial WSNs have been designed to only support static nodes and are not able to deal with mobility. In this paper, we show how a single-hop, multi-gateway IEEE 802.15.4e TSCH network architecture can tackle the mobility problem. We introduce the Virtual Grand Master (VGM) concept that moves the synchronization point from separated Backbone Border Routers (BBRs) towards the backbone network. With time synchronization of all BBRs, mobile nodes can roam from one BBR to another without time desynchronization. In addition to time synchronization, we introduce a mechanism to synchronize the schedules between BBRs to support fast handover of mobile nodes. 
\end{abstract}

\begin{IEEEkeywords}
Industrial WSN, 802.15.4e TSCH, IIoT.
\end{IEEEkeywords}

%
\IEEEpeerreviewmaketitle

\section{Introduction}
%
%
%
%
\label{introduction}

Under the umbrella of the Internet of Things (IoT) vision, a large number of objects are being connected to the Internet, sharing a huge amount of data that is consumed by applications. Nowadays IoT technologies are increasingly utilized in industrial settings. This Industrial IoT will help to improve the productivity, quality and robustness of the industrial processes.

Industrial WSNs are used to report measurement data to a central point. Such communication might be critical, requiring reliability up to 99\% in a harsh industrial environment. Moreover, latency determinism and low power usage is required too. In order to meet these requirements, Time Synchronized Channel Hopping (TSCH) became the main MAC technique in industrial standards like WirelessHART \cite{wirelesshart}. The new amendment of the IEEE 802.15.4-2011 \cite{802.15.4ieee2011ieee} standard, 802.15.4e \cite{802.15.4e}, introduces TSCH mode. In addition to this, the IETF working group of 6TiSCH is investigating end-to-end IPv6 connectivity over IEEE 802.15.4e TSCH \cite{rfc8180}. 

Both, WirelessHART and IEEE 802.15.4e, are foreseen to be used for static networks only. Mobility in IEEE 802.15.4e creates problem for time and schedule synchronization between nodes. Frequent resynchronizations are both power and time consuming processes. In addition to time synchronization, schedule needs to be updated too.

In this paper, the proposed architectures, namely a single-hop multi-gateway 802.15.4e TSCH network is designed and evaluated. This novel architecture combines the advantages of TSCH with support for mobility. More concretely, the following contributions are made. We present a single-hop 802.15.4e network architecture with multiple 802.15.4e gateways and mobile nodes. We introduce the network time grand master concept that makes the network to remain fully time synchronized, enabling the node handover without time desynchronization. In addition, the scheduling scheme to cope with mobility is presented together with testbed validation. 

The rest of the paper is organized as follows: in Section \ref{problem statement} the problem statement is discussed. Related work is presented in Section \ref{related work}, while Section \ref{system design} details the system design dealing with the network architecture, node synchronization and traffic management. Section \ref{results} presents the testbed evaluation results while, Section \ref{conclusion} concludes the paper.


 
\section{Problem Statement}
\label{problem statement}

Industrial environments are harsh environments characterized by metallic structures that obstruct the wireless communication. Different propagation effects makes these environments difficulties for wireless communication. Within such a challenging environment \cite{Karaagac}, we consider an automated warehouse system based on mobile transport vehicles, $'$shuttles$'$. These shuttles can move in 2D within the storing racks of a warehouse at reasonably high speed up to 3 m/s. 

Considering dense storage of goods, a multi-AP networking system needs to be considered to cover the whole racking system. Due to the relatively high speed of the 2D shuttles, the use of a multi-AP 802.11 network might result in frequent handovers that can go up to some hundreds of ms \cite{haxhibeqiri2016wireless}. This will increase the overall communication latency, or even worse, break the communication link for some time. In addition, 802.11's CMSA/CA mechanism is unable to offer bounded latencies. Moreover, shuttles always have to be connected to a central server via a reliable and real-time wireless communication link in order to provide timely status updates and to receive order assignments. 

Considering all of the aforementioned challenges, the optimal network solution for such a case needs to provide a handover latency that is lower than 1s, capacity requirements up to several bps, and the possibility to serve up to 100 shuttles at a time. The data packets will be 8 to 128 bytes with a maximal communication frequency of 1Hz in uplink. In addition to frequent uplink monitoring data, infrequent transmission of picking orders (downlink data) can happen. The reliability is targeted to be as high as 99\% in a controlled industrial environment. 

\section{Related Works}
\label{related work}

A number of studies treats different aspects of IEEE 802.15.4e TSCH networks such as network synchronization, network formation, node mobility support, traffic scheduling and power consumption. Here, we will only present works related to node mobility support and industrial applications.

In \cite{NovelRouting} a novel routing solution in TSCH networks with mobile nodes is presented. The network is composed by static nodes, called anchor nodes, and mobile nodes. Routing between anchor nodes is done using RPL while anchors' positions are known to the mobile nodes. Every mobile node estimates its distance to each of the anchors that minimizes the number of expected transmissions towards the sink. 

In \cite{MobilityAware} a mobility-aware TSCH framework is proposed that accelerates the node association process. For that MTSCH uses ACK messages as passive beacons. At the end of each slot frame, a node will send a group ACK on a specific channel offset to acknowledge all received packets from all of its neighbours together with information when the node will listen for any mobile nodes. Similarly, in \cite{ocari} the slot frame is divided in two parts: a CSMA/CA part and a TDMA part. For communication between mobile nodes and static ones, a CSMA/CA based scheme is used, while the communication between static nodes uses fixed time slots. 

In \cite{ldpq} authors present a TDMA MAC approach based on low power listening (LPL) and distributed queuing (DQ). The network is a single-hop network with multiple mobile nodes. The communication process is divided into two phases: a network synchronization phase and a data communication phase. During the network synchronization phase, a coordinator transmits wake-up packets with a certain frequency and using a specific channel. Once the mobile node receives such a wake-up packet, it synchronizes to the network, while the data communication period uses a time-fixed frame structure. 

\section{System Design}
\label{system design}

For our use case, we target a single-hop multi-gateway IEEE 802.15.4e TSCH network architecture. In order for such an architecture to enable seamless mobility and guaranteed deterministic communication, two major improvements are needed. First, mobile nodes need to be able to roam from one 802.15.4e gateway (or access point) to another without time desynchronization. Second, the mobile node should be able to maintain its schedule too. This new architecture and the study of both aspects is the main focus of this paper.

\subsection{Network architecture}

\begin{figure}[!t]
\centering
\includegraphics[width=3in]{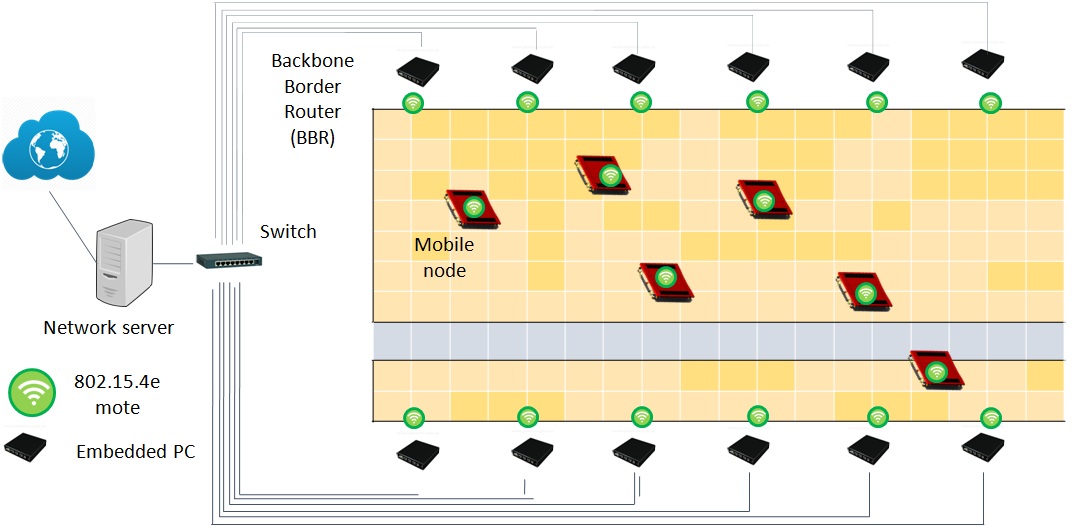}
\caption{Single hop network architecture for 2D shuttle systems.}
\label{Fig3}
\end{figure}

The key components of the network architecture are shown in Figure \ref{Fig3}. It involves the mobile nodes, the Backbone Border Routers (BBRs) and the Network Server (NS). All BBRs are connected via a wired backbone network to the Network Server (NS), while mobile nodes only use single hop wireless links towards the backbone, with the BBRs serving as access points. The NS, acts as the virtual root of this Low power Lossy Network (LLN), enabling time synchronization between the BBRs and performing traffic management. It ensures that all BBRs are listening at the same time to the same channel, that the schedules of all mobile nodes are known by all BBRs and that both uplink and downlink traffic are handled properly.  

\subsubsection{Backbone Border Routers (BBRs)}

The BBRs act as access points for the IEEE 802.15.4e mobile nodes. They serve as DODAG roots for the mobile nodes and broadcast the same IPv6 prefix to nodes that want to join the network. From the mobile node perspective, all BBRs will be seen as a single access point towards the backbone network regardless of their position in the network. Apart from using the same network address, BBRs need to be time synchronized in order to offer smooth handovers for the mobile nodes. This time synchronization between BBRs is done via the backbone network. Another enabler for seamless mobility is the traffic scheduling inside the LLN. To decrease the amount of signaling traffic for routing (such as Neighbor Discovery traffic) inside the LLN network, mobile nodes will send layer 3 unicast packets encapsulated in layer 2 broadcast packets. BBRs within range will receive the broadcast packets and will route them to the NS. Downlink traffic will be scheduled from the NS as unicast packets, by selecting the $'$best$'$ BBR for the downlink stream.

\subsubsection{Network Server (NS)}

The NS acts as a Virtual Grand Master (VGM) for the whole network. Based on its time, the BBRs and mobile nodes are being synchronized. In addition to network time synchronization, the NS performs deduplication of upstream packets from multiple BBRs and selects the best BBR for the downstream communication. This selection can be based on the RSSI or SNR values of the last couple of upstream packets. This information is collected by BBRs and is sent to the NS. The NS keeps track of the last BBRs via which the mobile node was reachable. The last function performed by the NS is the management and update of the BBRs' TSCH schedule. The mobile node will negotiate the schedule with the NS that will install the schedule in all BBRs.

\subsection{Synchronization of the backbone and TSCH network}

Mobile nodes would desynchronize every time they move from the coverage zone of one BBR to another. Every resynchronize will result in long periods without connectivity. To improve mobility handling and to reduce the synchronization time, we therefore propose BBRs to be time synchronized between each other by means of the NS. Once all BBRs are time synchronized, the mobile node does not need to perform reresynchronization. This requires a mapping between the time mechanism used in the 802.15.4e TSCH network and the one in the backbone network.

\subsubsection{Virtual Grand Master Synchronization} 

At the backbone network side, the BBR$'$s network initialization time is taken as a time reference to calculate the ASN. This reference time is communicated to the BBRs by the Network Server. This makes it possible to add new BBRs to the network without resetting the whole network and loose synchronization. The network keeps calculating the elapsed Absolut Slot Number (ASN) and the $'$\textit{clock ticks}$'$ (as seen by the 802.15.4e node) based on this time reference. This information is communicated periodically to the BBRs by the Network Server. By moving the clock master of the mobile nodes from separate BBRs towards the backbone network, all nodes are now synchronized with a single time source. This is referred to as VGM synchronization.

\subsection{Handling of upstream and downstream data traffic} \label{traffic_management}

The NS is responsible for managing the data traffic coming from and going to the LLN network. Upstream traffic from a mobile node can be received by multiple BBRs simultaneously, as all BBRs are synchronized and will listen to the same channel at the same time (see the next subsection). These BBRs will forward the traffic to the Network Server, which has to perform deduplication in order not to send the packet multiple times to the final destination. To this end, the Network Server maintains a deduplication table, by saving the hash value of received packets. If a duplicate packet arrives, the calculated hash value will already be present in the table and the NS will discard it. Next to this, the NS also has a BBR table, with timely entries, that keeps track via which BBRs each of the mobile nodes in the network can be reached. This table is updated upon the reception of upstream traffic from BBRs. This table is used for routing downstream traffic.
 
\subsection{Schedule Organization} 
\label{scheduling_org}

\begin{figure}[!t]
\centering
\includegraphics[width=3.5in]{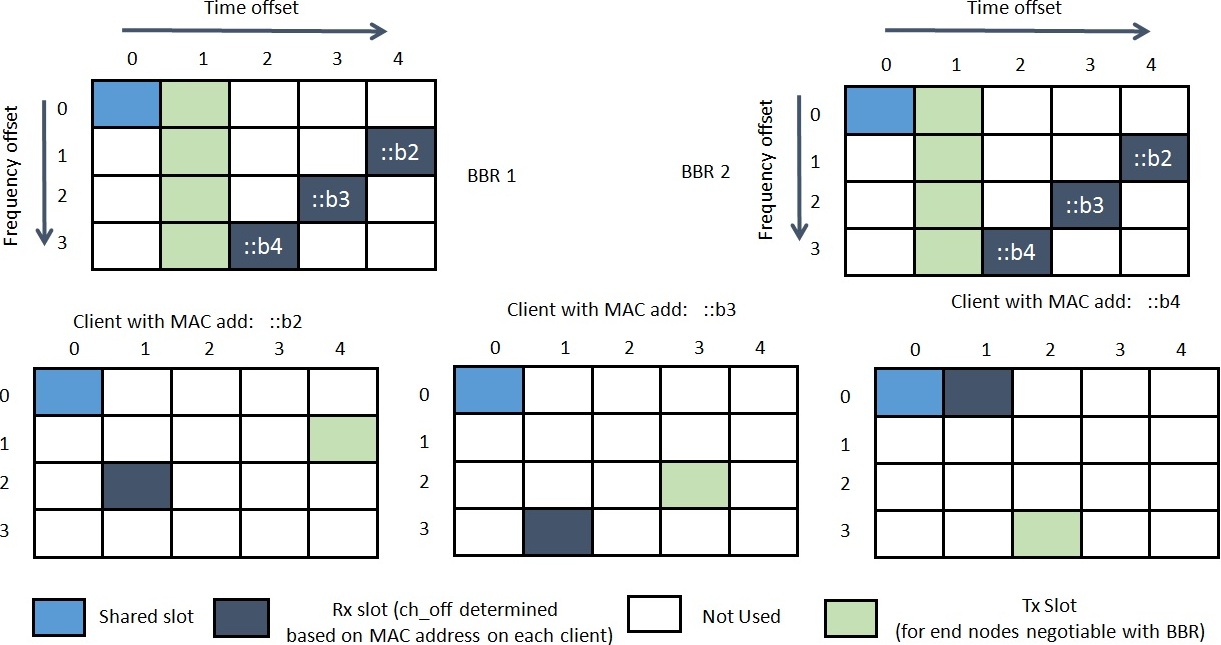}
\caption{Example of a schedule using shared downlink slot and negotiable uplink slot.}
\label{Fig4}
\end{figure}

Next to data traffic handling, the NS also plays a role in the scheduling process, resulting in the exchange of signaling traffic with the BBRs. We propose a number of upstream scheduling schemes, however, in this paper version we will describe only the most appropriate one. 

\textbf{Downstream scheme:}
For downstream traffic, we allocate a dedicated time slot in the slot frame. This time slot is determined by the NS and is communicated first to the BBRs and then to the mobile node itself by a BBR once it is synchronized with the network. The time slot for downlink traffic of different mobile nodes can only be shared in the time domain, as different channel offsets will be used for different mobile nodes. The actual channel offset used for downlink needs to be specific for each mobile node and to be determined dynamically. If there are more mobile nodes in the network than available channel offsets, additional downlink time slots can be added. For downstream traffic, spatial reuse is possible, as different BBRs can transmit at the same time to different end nodes by using different channels.

For upstream traffic we cannot use the same logic, i.e. using the same time slot and different channel offsets, as the communication is one to many. Due to the use of layer 2 broadcasting in the upstream direction, a mobile node must be able to communicate with multiple BBRs at the same time. 

\textbf{Upstream scheme}
The second scheme proposes a single transmit slot for each mobile node. This slot is negotiable with one of the BBRs, and once it is installed in the root BBR it will be communicated to all others via the Network Server. The negotiation of the schedule is done as in \cite{6top}. Hence, the mobile node can transmit its layer 2 broadcast upstream packets at a collision free time slot. The packet will be received by multiple BBRs as they are time and schedule synchronized. The broadcast nature of upstream traffic will imply the absence of MAC layer ACKs. However, the collision free transmission and duplication of packets (reception of the same packet by multiple BBRs) will boost reliability. An example of this scheme is shown in Figure \ref{Fig4}. For large scale deployments, this solution will increase the length of the slot frame, keeping the collision probability in upstream practically zero.

\section{Results}
\label{results}

In order to validate the proposed solution, we performed tests in the w-iLab.2 testbed, which is a generic wireless testbed in a pseudo-shielded environment. The first set of tests validates the synchronization between the VGM, the BBRs and the mobile nodes. The second set of tests validates the low latency handover of a mobile node from one BBR to another. The last set of tests assesses the impact of the different schedules we described in Section \ref{scheduling_org} on latency and packet losses. For all measurements in the next subsections, Zolertia Remotes rev B1 \cite{zol} have been used as IEEE 802.15.4e nodes. The OpenWSN \cite{openwsn} implementation of the IEEE 802.15.4e TSCH mode has been used as a starting point for the development of the system. In this version of the paper we will include only smooth handover results.

\subsection{Handover Latency}

In a network where all BBRs are fully time synchronized between each other the handover of mobile nodes should exhibit a low latency. We conducted a number of experiments with different traffic frequencies in the w-iLab.2 testbed. We used two static nodes with Zolertia Remotes rev B1 as BBRs and a mobile node which was roaming around. The distance between the BBRs was 35m while the length of the trajectory was 52m in a straight line. Both BBRs are connected to the backbone network where the Network Server runs the PTP server for network time synchronization. The transmit powers of the BBRs and the mobile node were 3dBm. In order to decrease the intermediate coverage zones we added a fixed attenuation of 20 dB to antenna ports of the BBRs. This resulted in a coverage zone of each of the BBRs of $\sim$25m. We generated an 80 byte packet every 300ms and 1sec, respectively, from the Network Server towards the mobile node. The RTT latency was measured as well as packet losses.

\begin{figure}[!t]
\centering
\subfloat[]{\includegraphics[width=2.8in]{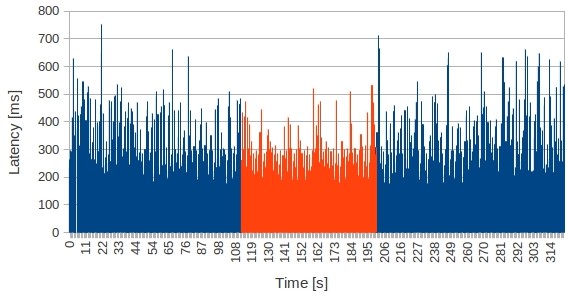}%
\label{fig81}}\\
\hfil
\subfloat[]{\includegraphics[width=2.8in]{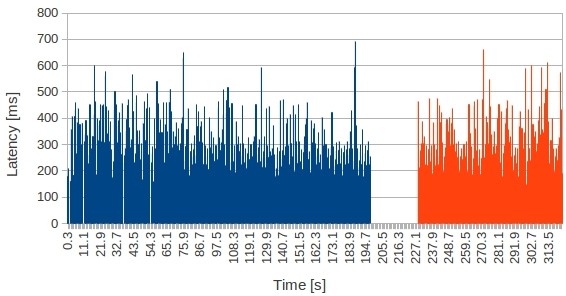}%
\label{fig82}}
\caption{The packet latency during handover for traffic frequency of 1 Hz. a) All BBRs are synchronized. The read part shows packets that are received by both BBRs. b) BBRs are not synchronized. The read part shows packets that are received by second BBR.}
\label{fig8}
\end{figure}

In Figure \ref{fig81}, the latency over time is shown for synchronized case is shown. As it can be seen, there is no packet loss when the handover happens. The red part of the latency graph shows the latency of packets that were received by both BBRs in the intermediate coverage zone. When the BBRs are not time and schedule synchronized then handover latencies can be as high as tens of seconds. Such a case is shown in Figure \ref{fig82}. When the mobile node moves out of the coverage zone of the first BBR, it needs time to get synchronized with the new BBR. During that time, all communication is lost.

\section{Conclusions}
\label{conclusion}

By adopting time synchronized based MAC schemes it has become possible to deploy WSNs that meet strict latency requirements for industrial settings. However, their usage has been limited to statically deployed nodes. 

This paper has introduced a real-life industrial use case from the logistics domain that requires both the determinism as well as support for mobility. As a solution, we propose a novel single-hop multi-gateway IEEE 802.15.4e TSCH network architecture where all the gateways, called BBRs, are both time and schedule synchronized by means of a Virtual Grand Master residing at the Network Server. We have addressed the synchronization and scheduling challenges that come with this architecture and have shown that this architecture is indeed able to improve the handover latency and decrease the communication outage time due to desynchronization. 

The NS performs traffic management functions: packet deduplication in the uplink, while it selects the best BBR for downlink communication. For the schedule synchronization we have presented the scheduling scheme that offers collision free time slots for uplink and downlink communication and it is maintained for end nodes throughout the network. We may conclude that the proposed architecture is viable and able to deal with mobility while offering communication guarantees by means of the installed schedules. Of course, some capacity must be sacrificed in order to support seamless handovers. The proposed concept can be applied in a broader context, for instance, it would be worth exploring how the same principles can be extended to higher bit rate radios such as Wi-Fi or to e.g. large-scale Ultra wide-band (UWB) localization networks.


%



\section*{Acknowledgment}

HYCOWARE is a project realized in collaboration with imec. Project partners are Egemin, Aucxis and Intation, with project support from VLAIO (Flanders Innovation and Entrepreneurship).

\ifCLASSOPTIONcaptionsoff
  \newpage
\fi



\bibliographystyle{IEEEtran}
\bibliography{IEEEexample.bib}
%



%




\end{document}